\documentclass{elsarticle}

\usepackage{color}
\usepackage{url}
\usepackage{draftwatermark}
\SetWatermarkLightness{0.7}
\SetWatermarkScale{1.3}
\SetWatermarkText{arXiv:submit/1276301 [cs.IR] 10 Jun 2015}

\begin{document}
\title{Increasing loyalty using predictive modeling in \\  Business-to-Business Telecommunication}
\author{Patrick Luciano$^{1}$, Isma\"\i l Reba\"\i$^{1}$, Vincent Lemaire$^{2}$}
\address{(1) Orange Business Services, 27 rue M\'ed\'eric, 75840 Paris, France \\
(2): Orange Labs, 2 avenue Pierre Marzin, 22300 Lannion, France}

\begin{abstract}
Customer Relationship Management (CRM) is a key element of modern marketing strategies. One of the most practical way to build useful knowledge on customers in a CRM system to produce scores to forecast churn behavior, propensity to subscribe to a new service... In AMEA zone (Asia, Middle East and Africa zone), the context of fierce competition may represent a higher percentage, and particularly in B2B market. But by contrast, to our knowledge, no scientific papers were dedicated and published to detail the way to improve loyalty in B2B Telco market. If we can assume at low segments similarities between B2B and B2C, some research is required in order to model B2B user behavior versus B2C behavior. This problematic stands actual as ``Bring Your own Device'' (BYOD) becomes more and more trendy. Answering business requirements, our team applied some B2C predictive tools with adapting them to B2B in an AMEA country Orange affiliate.

\end{abstract}

\begin{keyword}
value management\sep predictive modeling\sep loyalty\sep CRM\sep marketing\sep B2B\sep Telco.
\end{keyword}

\maketitle


\section{Introduction - Marketing objectives: the interest of CRM tools to improve customer retention}

Scholars have questioned the effectiveness of several customer relationship management strategies. The differential effects of customer relationship perceptions and relationship marketing instruments on customer retention and customer share development over time have been investigated. Verhoef \cite{Verhoef:2003} has showed that loyalty programs that provide economic incentives positively affect customer retention.

\subsection{The role of marketing on a business's performance}

One of the main objectives of the marketing function is to increase the value generated by actual and future customers. The customer is key in the company strategy. ``The Consumer, not the Company is in the middle'' \cite{Keith:1960}.

Literature has shown that a market oriented organization supports the company's performance. Kohli \& Jaworski \cite{Kohli:1990} have developed a framework for understanding the implementation of the marketing concept.

Marketing academicians and practitioners have been observing that business performance is affected by market orientation \cite{Kirca:2005}. Narver \& Slater \cite{Narver:1990} developed a valid measure of market orientation. They analyzed its effect on a business's profitability. Using a sample of 140 business units consisting of commodity products businesses and no commodity businesses, they found a substantial positive effect of a market orientation on the profitability.

\subsection{Value management}

Marketing theory and practice has become more and more customer centered. Managers have increased their emphasis on long-term client relationships because the length of a customer's tenure is assumed to be related to long-run company revenues and profitability \cite{Gupta:2004}.

\subsection{The role and interest of CRM tools}

Customer Relationship Management (CRM) is organized according to the customer lifetime duration because lifetime with a firm generally is not perpetual. Consumers may be dissatisfied and find better value elsewhere \cite{Oliver:1999}.

\subsection{The effect of loyalty programs / management}

Meyer-Waarden \cite{Meyer-Waarden:2006,Kumar:2004} examines the impact of loyalty programs on customer lifetime duration. He shows that loyalty schemes have positive effects on customer lifetimes.

\subsection{Churn is a key indicator of customer loyalty}

The churn, or customer attrition, is defined as the number of disconnections for the period divided by the average base for the period where the average base is the sum of the opening base at the beginning of the period and the closing base at the end of the period divided by 2. We distinguish the commercial churn from the financial churn. The churn is a key indicator, mainly in the financial sector (banking and insurance) and the telecommunication industry.


\section{Forecasting / Predicting the churn}

\subsection{Churn Prediction}

As described in the first section of this paper. The churn definition is not the same in postpaid and prepaid context. We focus here on the former: the propensity of customers to switch provider in postpaid mode even if the methodology described share many aspects with prepaid prediction.

\subsection{Forecasting aspects}
\label{Forecastingaspects}

Given a database, the most common task in data analysis is to find the relationships between a set of input or explanatory variables (X) and one target variable (Y). This knowledge extraction often goes through the building of a model (f) which describes these relationships. For all the instances of the database, a probabilistic model allows, face to a classification problem and given the values of the explanatory variables, the estimation of the probabilities of occurrence of each class target. The model learns the relation f: X $\rightarrow$ Y.

\begin{figure}[h!]
\begin{centering}
\includegraphics[width=1.0\textwidth]{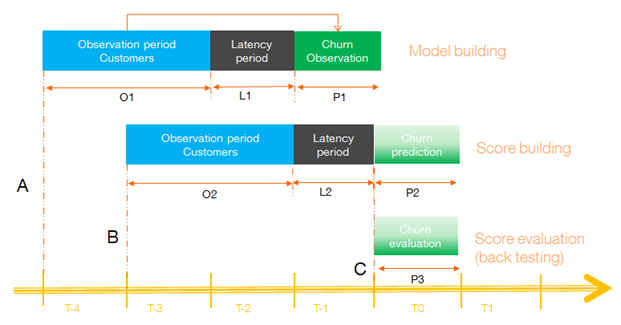}
\par\end{centering}
\caption{\label{time} Time periods definition}
\end{figure}

This model is trained based on time windows and applied on a sliding window (see Figure \ref{time}). Different time periods definitions exist:

\begin{itemize}[$\bullet$]
\item Observation period: period used to observe the customer behavior and find the right explanatory variables (see Section \ref{explanatoryvariables}) to infer their future behavior
\item Latency period: period required to compute the score and required by the marketing services to be able to launch the campaign based on the model scores
\item Churn observation: period on which the churn is observed and used to train the model
\item Churn prediction: period on which one wan to deploy the scoring model
\item Churn evaluation: period on which the churn is observed and evaluate the prediction performances of the model applied on the ``churn prediction'' period (back testing)
\end{itemize}

\newpage

For example:
\begin{itemize}[$\bullet$]
\item Step 1: at time ``A'' using an observation period (months T-4 and T-3) ``O1'' and churn observation ``P1'' a scoring model is trained. In this step one are in a classical data mining process [REF] in which during the modeling step the data are split in a train part and a test part to evaluate the quality of the model.
\item Step 2: then this model is deployed a time ``B'' using an observation period (months T-3 and T-2) ``O2'' to predict the churn on the period ``P2''
\item Step 3: later a time C when the true information about the churn is available the performances of the model deployed at the step 2 are evaluated (back testing).
\end{itemize}

Note 1: Usually the size of O1, 02 and L1, L2 and P1, P2, P3 are the same respectively. More L is wide more the problem is difficult.

Note 2: A difficult point is to obtain the same performances on the test dataset in the step 1 and the performances in the step 3 due to many aspects such as for example non stationary environment.

Note 3: In step 1 one needs customers which are present in ``O1'' and remain present until the beginning of ``P1''.

\subsection{Data sources}
\label{datasources}

 \begin{figure}[h!]
\begin{centering}
\includegraphics[width=1.0\textwidth]{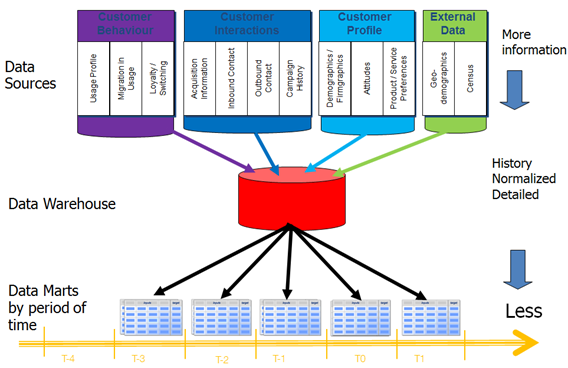}
\par\end{centering}
\caption{\label{data}Data sources and data mart by period of time}
\end{figure}

The score is obtained using the quantitative information and qualitative information available on the user from different sources such as socio-demo, behavior of purchase, preceding answers, ... This sources filled a data warehouse (the red module in Figure \ref{data}) which contains tables connected in relational scheme where for examples a ``customer table'' (the root table) is connected with others ``secondary'' tables (as Usages, Products, ...).

The root table is related to the marketing domain. For customer data analysis, this table contains all the fields directly connected to the customer, such as his name or address. The secondary tables have a 1-N relationship with the primary table. Each instance of the primary table may be related to a variable number of instances of a secondary table. For telecommunication data for example, the secondary tables contains the list of services, of usages of these services, the call details.

The used of this different sources, different ``kinds'' of databases, requires the use of ``model'' able to deal a variety of difficulties: heterogeneous data (numerical and categorical variables), noisy data, unbalanced distributions of predictive variables (very few non-zero values) and target values (only very small percent of the examples belong to the positive class), sparse variables and many missing values.

\subsection{Explanatory variables}
\label{explanatoryvariables}

A score\footnote{Technique of hierarchisation of the users which makes it possible to evaluate by a note the probability that a user answers a request or belongs to the required target within the framework of a direct marketing campaign}  (the output of the model described in section \ref{Forecastingaspects}) is an evaluation for the target variable, here the churn, to explain. The score is computed using customer records represented by a number of variables or features: the explanatory variables. Scores are then used by the information system (IS), for example, to personalize the customer relationship. The explanatory variables come from (at least) three origins:
\begin{itemize}
\item Native variables from the IS: these variables are directly present in the root table (see Section \ref{datasources}) and can be extracted for example by ``Get'' instruction
\item Aggregates from experts: these aggregates are variables that are elaborated by a formula using the root table and secondary table(s), the formula is given by an expert of the domain who has the expert knowledge that this aggregate is useful for a given problem. For example, to build the number of usages of each service per weekday for all customers, one single language expression needs to be specified, with the use of the ``Count'' operator on the secondary table ``Usage'' with two operands ``WeekDay(Date)'' and ``Label(ServiceId)''
\item Aggregates automatically constructed: these aggregates are also variables that are elaborated by a formula but without expert knowledge. The idea is not to elaborated randomly potential aggregates but  to use a prior distribution  over  the  space  of  all  variables  that  can  be  constructed in  the  basis  of  a  description  of  a multi-table schema and a set of construction rules. This allows building tens of thousands of features to create a rich data representation space. This point relies on domain knowledge to construct new potentially informative variables \cite{Boulle:2014}.
\end{itemize}

Whatever is the origin a particular explanatory variable, it has to be robust versus the time: i.e. this variable should be:
\begin{itemize}
\item useful when training the model (in Step 1 in Figure \ref{time}) and useful when using the model (in Step 3 in Figure \ref{time})
\item exhibits a good robustness when training the model and a good robustness when using the model (between Step 1 and Step 3 in Figure \ref{time})
Note: we call here robustness the performances ratio between the ``training dataset'' and the ``test dataset'' during the training process OR the performances between ``test dataset'' and the ``back testing dataset'' later. We are interesting in a ratio close to 1.
\end{itemize}

\subsection{Automation and Volume}

A good tool should be able to:
\begin{itemize}
\item extracts automatically a large number of features from a relational database (relies on domain knowledge to construct new potentially informative variables)
\item selects a subset of informative variables and instances \cite{Feraud:2010}
\item builds the predictive model for datasets having a very large number of input variables (thousands) and instances (hundreds of thousands)
\item allows a robust detection of the most predictive variables
\item and efficiently builds an accurate classifier (the model).
\end{itemize}

A key requirement is the complete automation of the whole process.

\subsection{Evaluation of the model quality}

\begin{figure}[h!]
\begin{centering}
\includegraphics[width=0.6\textwidth]{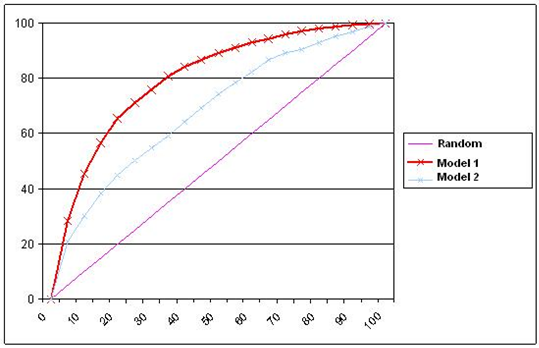}
\par\end{centering}
\caption{\label{lift} Lift curve: an example}
\end{figure}

There is value in a CRM system to evaluate the ``propensity'' of customers to churn. Therefore, tools producing scores are more usable that tools producing only binary classification results. The tool is then asked to provide a score (a discriminant value or a posterior probability (P(Churn=1 |X)). The quality of the model is then judged by the AUC\footnote{ The AUC is a standard metric in classification see : {\url{http://en.wikipedia.org/wiki/Receiver_operating_characteristic}}}  (the area under the ROC curve), which is a scalar between 0.5 and 1.0 (0.5 corresponds to a random model and 1.0 corresponds to the `oracle' model) and the lift curve (see below Figure \ref{lift}).

The AUC is related to the lift curve. On the horizontal axis there is the percentage of the total population (of your marketing campaign) and on the vertical axis there is the percentage of the target population, for example the churn population. In the figure here we have two lift curves. The red curve has been obtained using the model 1 and the blue one has been obtained using the model 2. If we use the model 1 and we send an email to 20\% of the total population, we will touch only 40\% percent of the target. But if we use the model 2 we will touch 60\% percent of the population. Therefore a good lift allows a good campaign.

\subsection{Life cycle of a score}

The 'life' of a score is constituted by 4 main step: the request, the modeling, the deployment and the supervision.
\begin{enumerate}[$\bullet$]
\item Request Step:
\begin{itemize}
\item Demand: A business unit wants the computation of a score on Customer
\item Instruction of the demand: the score team studies its advisability and feasibility
\item Validation and prioritization of the demand
\item Specification
\end{itemize}
\item	Modeling Step as every data mining process, we have
\begin{itemize}
\item data acquisition
\item modeling (the training of the model)
\item validation
\item economic validation (do the results seem to be sufficient for a marketing campaign ?)
\end{itemize}
\item	Deployment Step:
\begin{itemize}
\item cabling
\item production
\item validation
\end{itemize}
\item Supervision Step
\end{enumerate}

\subsection{Key points in B2B - Individual interpretation}

\subsubsection{Common point with B2C}

In churn prediction in B2B and B2C, from the different points listed above in this section we may write a list or requirements:
\begin{enumerate}[$\bullet$]
\item Data constraints
\begin{itemize}
\item Heterogeneous
\item Missing values
\item Multiple classes
\item Heavily unbalanced distributions
\end{itemize}
\item Training requirements
\begin{itemize}
\item Fast data preparation and modeling
\item	Variable construction
\end{itemize}
\item Model requirements
\begin{itemize}
\item Reliable
\item Accurate
\item Parsimonious (few variables)
\item Robust
\item Understandable
\end{itemize}
\item Deployment requirement
\begin{itemize}
\item Fast deployment
\item Close to real time classification 
\end{itemize}
\end{enumerate}

To achieve this points the Khiops\footnote{Available on {\url{www.khiops.com}}}  software was used. Khiops is a data preparation and modeling tool for supervised and unsupervised learning. It exploits non parametric models to the predictive importance of input variables in the supervised case. These evaluations are performed by the mean of discretization models in the numerical case and value grouping models in the categorical case, which correspond to the search for an efficient data representation owing to variable recoding. The tool also produces a scoring model for supervised learning tasks, according to a naive Bayes approach, with variable selection and model averaging.

The tool is designed for the management of large datasets, with hundreds of thousands of instances and tens of thousands of variables, and was successfully evaluated in international data mining challenges. This tool contains also a framework aiming at automating variable construction for supervised classification \cite{Boulle:2014}.

In B2B as for B2C the purpose is to improve customer experience: leverage each interaction whatever the channel is, to identify the customer and provide him a differentiated treatment. A key point is therefore the interpretation of the results obtained with the scoring model. The level of ``interpretation'' could not be the same in B2B and B2C as described in the next section.

\subsubsection{Key point in B2B}

The scores are not always directly usable. For example if a scoring model identifies a customer interested in churning, the score does not say anything on the action needed to avoid his cancellation. To prevent him from churning, his reasons have to be identified.

The reasons of the churn could be given at three levels of granularity:
\begin{itemize}
\item Global: the reasons of the churn are given in average for the entire population
\item Local: the reasons of the churn are given in average for a segment of the population
\item Individually: the reasons of the churn are given customer per customer.
\end{itemize}

In B2B because of the volume (i.e. the number of customers concerned by the future retention campaign to launch) the reasons of the churn are mainly described at the global level. This is done often by simply giving the importance of the explanatory variables placed at the input of the scoring model for the global population. The objective of the measure of the importance of the input variables is four-fold: improve the prediction performance of the predictors, provide faster and more cost-effective predictors, allow a better understanding of the underlying process that generated the data and simplify the understanding of the results delivered by the model. For the Khiops software this is done by giving the weights of the variables. These weights allow to say that in average a given variable is the most important to predict the churn.

In B2C since the volume is lowest it is possible to go down from one or two levels. For the individual level one will be interested by: a methodology which would, for every customer, (i) identify the importance of the explanatory variables; (ii) identify the position of the values of these explanatory variables; and (iii) propose an action in order to change the customer response. The first point should answer to ``Why THIS customer is predicted to go to churn'' and the third one ``How is it possible to keep THIS customer''. The reader interested by this methodology may read \cite{Lemaire:2009}.

The Khiops software has therefore a complementary module, the add-on ``Khiops Interpretation'' that allows to obtain this couple of information. Understanding the score given by a model for a particular instance can for example lead to an immediate decision in a Customer Relationship Management (CRM) system.


\section{Use Case: Churn prediction in B2B}

\subsection{Preliminaries and context}

\subsubsection{Definitions used}
In this section, and before modelling churn and using customers' value analytics, we fixed some preliminaries:
\begin{enumerate}[$\bullet$]
\item A customer is: a person using a card and identified by its MSISDN (Mobile Station Integrated Services Digital Network) commonly the GSM phone number. We will focus on his mobile consumption, for problem simplification. We distinguish between his in net and off net calls, as well as his international calls. All usages will be included: voice, SMS and data. As we are studying the B2B market, we will include all segments, i.e. low SoHo (Small office, Home office) segment, mid SME (Small \& Medium Enterprise) segment, as well as high LA (Large Account) segment.
\item A churner is someone (end-user) who: has disconnected from one month to another. This is the TARGET VARIABLE. A pre-paid churner is a user who did not refill in the previous two months. A post-paid churner is the one who stops the contract. We excluded bad debts as they do not have the same behavior as potential churners.
\item A churner is someone (account) who: has a decrease of at least 25\% of the total number of MSISDNs compared to previous month.
\end{enumerate}

Our aim is to identify the end-users presenting the risk to leave the company, in order to retain them. What should be underlined is that in B2C churn modelling concerns the end-user. However, in B2B, we need to link the end-user to the account that identifies the company of the end-user.

Another declination of the churn modelling concerns pre-paid offers, post-paid offers and hybrid offers. In order to simplify the problem, we did not make difference for churn forecasting between the type of the offer. One of the reasons is that pure post-paid and post-paid part of hybrid offers correspond to B2B usage and pre-paid is a more B2C usage.

\subsubsection{Explanatory Analysis of the used data}
\label{nbvars}

\begin{enumerate}[$\bullet$]
\item What to fetch: The objective is to identify correlations between different factors (past activity, business demographics, etc.) and future activity, i.e. to estimate a model of customer activity.
\end{enumerate}

The modeling will focus on customers' churn probability. It is equally critical to prioritize anti-churn actions based on customer revenues:
\begin{itemize}
\item Statistical analysis and forecasts
\item Draw insights from the model to support next action to take
\item Document model and model findings
\item Produce scoring for each customer
\end{itemize}

\begin{enumerate}[$\bullet$]
\item How to fetch: The data collected had to be analyzed, cleaned and modeled.  
\end{enumerate}

\begin{figure}[h!]
\begin{centering}
\includegraphics[width=1.0\textwidth]{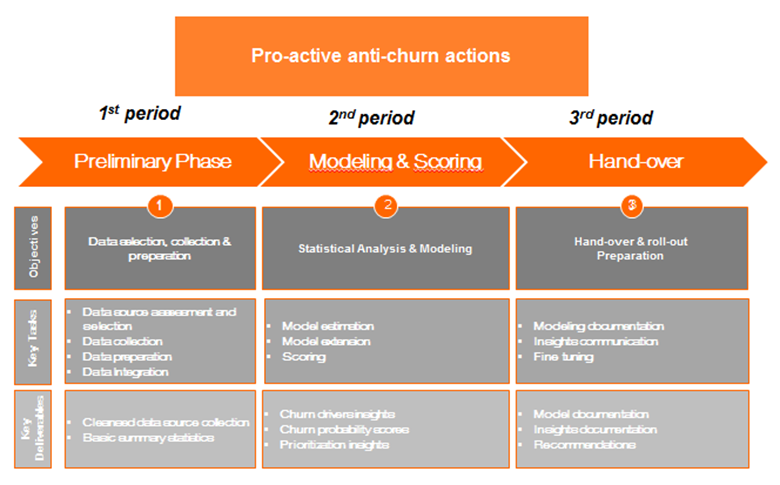}
\par\end{centering}
\caption{\label{methodo} Methodology}
\end{figure}

Initial received data summarized ~500 information per MSISDN (e.g. voice, SMS, sales representative in charge of the account etc.).
We followed a methodology based on a three steps close to the classical data mining process (see Figure \ref{methodo}).

\begin{figure}[h!]
\begin{centering}
\includegraphics[width=0.8\textwidth]{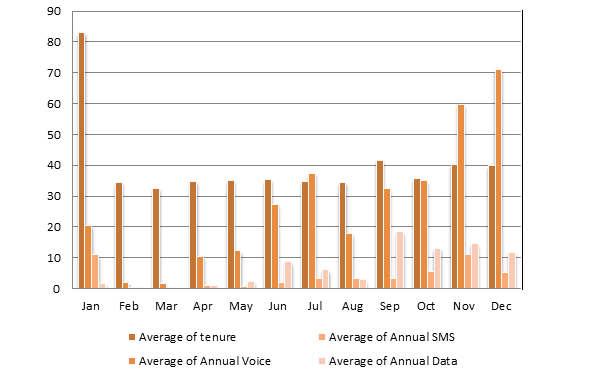}
\par\end{centering}
\caption{\label{insights} Example of an insight}
\end{figure}

In parallel of model building, we tried to get some insights from data. We analyzed different variables and tried to find correlations with churn. In the figure \ref{insights}, we fetched for any relationship between churn and tenure, Voice, SMS or data usage
As it can be seen, no obvious relationship was found.

\subsubsection{Amount of explanatory variables used}
 
During building models, we applied backward elimination variable approach \cite{BoulleJMLR07} reducing the number of variables in iterative steps according to models efficiency. If the model performance was similar, then we would drop the variables as explained:
\begin{enumerate}[$\bullet$]
\item Iteration 1: lagged variables out of model building history window scope.
\item Iteration 2: all remained variables taken for modelling, and then we decrease the number of variables until getting to an optimal number. We removed variables with (1) small weights\footnote{To simplify: the weights are the importance of the variables when they are together at the input of the scoring model and the levels are univariate importance of each variable when using only one of them, see \cite{BoulleJMLR07} for more details.} calculated by Khiops, (2) low levels calculated by Khiops.
\item Iteration 3: intermediate variables : dropped if accuracy not proved on models built on it.
\end{enumerate}

\subsection{Experimental results:  Models and scores delivery}

The model aims to predict churners (according to the definition) at end-user level and not at account level. The principle of the approach is to classify the identifiers of the end-users, by decreasing probability to churn. We hope by that, to find in the first sorted end-users the highest potential churners in order to focus retention actions toward them. This is justified by operational reasons, as we can nott afford to contact all customer data base each month to retain them. 
 
We found after different iterations with the ``classical'' data mining process \cite{Fayyad:1996} that the optimal number of explanatory variables (see Section \ref{nbvars}, with accurate models is obtained by four months of historical data (O1=2 months in the Figure \ref{time}), with a two months horizon forecasting (P1=2 months in the Figure \ref{time}) and no latency period  (L1=0 in the Figure \ref{time}).
 
The variables retained belong to the following classes:
\begin{enumerate}[$\bullet$]
\item Active Mobile: MSISDN, internal identifier, Code of the account, offer, activation date, last status date, segment, section, activation type
\item No active Mobile: MSISDN, internal identifier, Code of the account, churn type, segment, offer, activation date, deactivation date, reason
\item Trouble Tickets: internal identifier, section, ticket number, service domain, status
\item Yearly Customer Revenue: internal identifier, section, data revenue, fixed revenue, mobile revenue
\item Usage: MSISDN, destination, service, charge
\item Mobile characteristics: MSISDN, internal identifier, Code of the account, Sales Representative Code, name, account manager, handset type, handset subtype, handset date
\item Total bill value: for each MSISDN
\end{enumerate}

The lift curve results of the models obtained was one of the result criterion used to qualify the quality of our models. In this study the model obtained an AUC sufficient to be judge at "deployable".

Once the probability to churn is calculated at the end-user level, we proposed to the marketing team different ways of operational converting the results in retention campaigns:
\begin{enumerate}
\item  taking the first $A$ end-users and sorting them by most frequent companies to which they belong.
\item taking the first $B$ end-users and sorting them by the highest revenue companies to which they belong
\item  for all the database, calculate for each company the average probability of churning of all end-users of the company.
\end{enumerate}
where the value of $A$, $B$ are parameters to be discussed with the marketing team.

\begin{figure}[h!]
\begin{centering}
\includegraphics[width=0.6\textwidth]{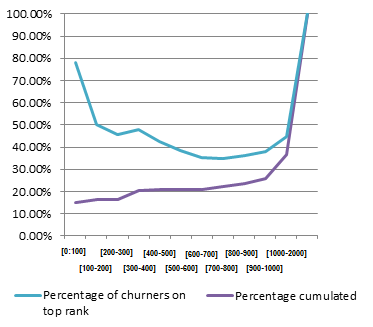}
\par\end{centering}
\caption{\label{percent}Percentage of churners in the ranked list}
\end{figure}

The model built for predicting churners detected 63\% of the real churners in the top 100 of the potential churners and 37\% in the top 2000, as shown in the Figure \ref{percent}

Moreover, thanks to Khiops interpretation, we could in addition to the probability of churning, get the most important variables to act on in order to decrease the churn probability. This allows giving for each end-user an operational lever to act on.

\begin{figure}[h!]
\begin{centering}
\includegraphics[width=1.0\textwidth]{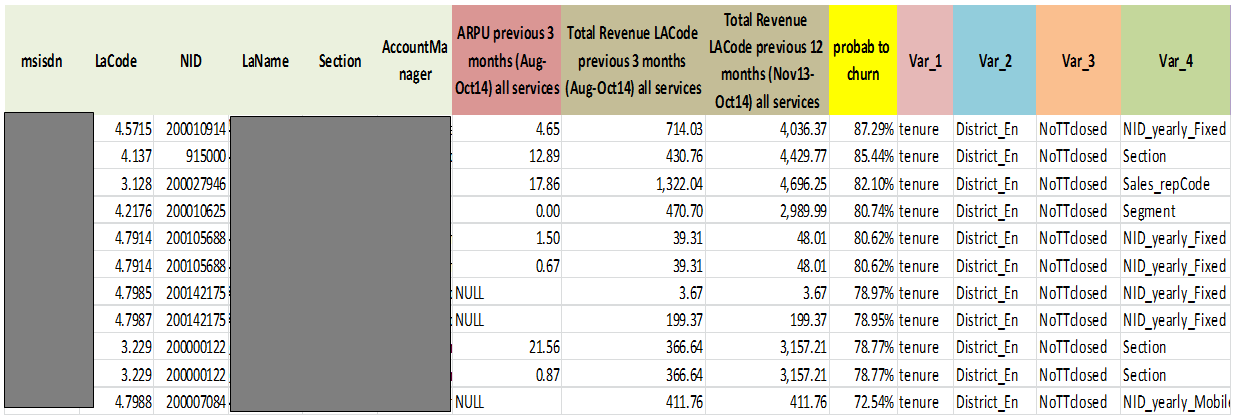}
\par\end{centering}
\caption{\label{How}``How'' indication \cite{Lemaire:2009}}
\end{figure}

In the table \ref{How}, the Var\_1 to Var\_4 are the reinforcement variables on which operational marketing can build the retention speech.

Using the scores and the "why" and "how" indications given respectively by Khiops and Khiops Interpretation the first proactive retention campaign targeted 1007 accounts, divided as the following: (i) Account Level - 27 accounts: Accounts with over 60\% of the account identified as risky, (ii) User Level - 980 accounts: Accounts having 59\% or less of the account's MSISDNs identified as risky.

The results were: (i) At MSISDN level, the churn dropped at 6\% vs 34\% (control group), (ii) At account level, the churn dropped at 4.5\% vs 80\% in the control group.

Note: To maintain the accuracy of the model at a convenient level, a periodic refresh of the model is required. It can be done each semester or after any important evolution in the offer or the market. As about recalculating the predictive scores, input variables refresh is a requirement. The predictive scores should be calculated at least every month (if not, bi-monthly). The model is a compilation of different sources as shown in the Figure \ref{data}. Refreshing the model is key for efficiency.

\section{Conclusion}

Following the CRISP-like methodology (CRISP stands for Cross Industry Standard Process for Data Mining) we used a scoring approach to improve loyalty in B2B. Different scenarios were tried to optimize the lift curve with as less as possible explanatory variables. We built different models playing with the historical window used for train, and the window used for test (using accuracy and AUC tests to choose the best model).  We then deployed the model, in order to provide operational marketing team with high probability churners. The results obtained, according that models have to be refreshed regularly, are very interesting in terms of accuracy and efficiency of the marketing campaign. More with help of variable importance and interpretation on the scores obtained ``customer per customer'', we obtained not only a ranking from a black box but  actionable actions.


\bibliographystyle{ijf}
\bibliography{mybib}
\end{document}